# Mentions of Prejudice in News Media – An International Comparison

David Rozado

Otago Polytechnic, New Zealand

david.rozado@op.ac.nz

ORCID: 0000-0001-6849-4746

## Abstract

Previous research has identified a post-2010 sharp increase of terms used to denounce prejudice (i.e. *racism*, *sexism*, *homophobia*, *Islamophobia*, *anti-Semitism*, etc.) in U.S. and U.K. news media content. Here, we extend previous analysis to an international sample of news media organizations. Thus, we quantify the prevalence of prejudice-denouncing terms and social justice associated terminology (*diversity*, *inclusion*, *equality*, etc.) in over 98 million news and opinion articles across 124 popular news media outlets from 36 countries representing 6 different world regions: English-speaking West, continental Europe, Latin America, sub-Saharan Africa, Persian Gulf region and Asia. We find that the post-2010 increasing prominence in news media of the studied terminology is not circumscribed to the U.S. and the U.K. but rather appears to be a mostly global phenomenon starting in the first half of the 2010s decade in pioneering countries yet largely prevalent around the globe post-2015. However, different world regions' news media emphasize distinct types of prejudice with varying degrees of intensity. We find no evidence of U.S. news media having been first in the world in increasing the frequency of prejudice coverage in their content. The large degree of temporal synchronicity with which the studied set of terms increased in news media across a vast majority of countries raises important questions about the root causes driving this phenomenon.

## Introduction

Previous research has identified a marked surge in the number of references to different prejudice types regarding ethnicity, gender, sexual or religious orientation in popular American news media outlets, both left- and right-leaning [1]. Follow-up research showed that these trends were not circumscribed to United States news media but that similar dynamics were also apparent in U.K. and Spanish media [2], [3]. Related trends in U.S. and U.K. news media usage of terminology often associated with social justice discourse such as *diversity*, *inclusion* or *equality* have also been documented [2], [4].

Critically, the existing literature has noted that the increasing number of references to prejudice in news media started prior to Donald Trump's 2015 emergence in the U.S. political arena when he announced his bid for the nomination of the Republican Party to the presidency of the United States. The trend also seemed to persist during at least the first few months following his departure from office [5]. As predicted by *agenda-setting* theory, the increasing prominence of terms signifying prejudice in U.S. news media was predictive of shifts in increasing concern among the American public about the severity of some, but not all, types of prejudice [1], [6], [7], [8].

The growing number of references to prejudice and social justice discourse is not circumscribed to news media. A content analysis of millions of published scholarly articles also showed an increasing frequency in the usage of such terms in the academic literature [9]. An important difference between news media and academic content was however apparent. While the trend in the usage of prejudice denouncing terms and social justice associated terminology in U.S. news media has been relatively flat since the 1970s until the early 2010s and it then explodes thereafter [1], [3], the trend in academic content had been growing up gradually since the 1980s [9], although an abrupt increase in the prevalence of this terminology is also apparent post-2010, but the rate of change appears milder than in news media.

Others have named the news media, academic and social dynamics delineated above as the *Great Awokening*. The term *woke* is an English adjective originally coined to denote awareness about prejudice and discrimination [10]. Although it was initially intended to signal positive connotations, over time it has also acquired negative undertones as a result of it being used by critics as a way of signaling what they perceive as exaggerated claims about the severity of prejudice and discrimination in modern Western societies [10]. Regardless of connotations, most observers would agree that the so-called *Great Awokening* can be characterized as a social phenomenon manifesting as increasing social and institutional focus on denouncing and combating presumed prejudice.

Renowned economists Tyler Cowen argued in an open editorial piece titled *Wokeism Will Rule the World* [11] that so-called *wokeness* is an American cultural export and that despite displaying some excesses, it will mostly have a beneficial impact in importing countries since injustice and discrimination are pervasive around the globe. The piece predicted that *wokeness* would spread worldwide and assumed that it began in the United States. Other authors have made similar assumptions regarding the U.S. origins of the phenomena [12]. Yet, empirical evidence that this is the case is lacking in the existing literature.

In previous works, we have operationalized the prevalence in news media of terminology often associated with the *Great Awokening* by longitudinally tracking the number of mentions of words used to denounce prejudice and terms often associated with social justice discourse. Here, we use similar methods to document the usage of such terminology in news media worldwide. We also attempt to characterize the emphasis of news media from different world regions on distinct types of prejudice and the correlation in news media worldwide between prejudice signifying words with negative connotations (*racism*, *sexism*, *homophobia*, etc.) and terms often associated with social justice discourse and positive connotations such as *diversity*, *equity*, *inclusion*, *fairness*, etc. Finally, we also try to elucidate whether the United States news media was the pioneer in increasing the usage of such terminology in their content.

This study does not directly resolve the question of the media's role as either a reflector or shaper of public opinion with respect to the topic of prejudice. However, it sheds light on the prevalence dynamics with which news media worldwide mention different types of prejudice, indicating the degree to which global media content is synchronized in highlighting issues of alleged discrimination. The research methodology also allows us to identify which forms of prejudice are most emphasized by news media in various countries and to track the changes in this emphasis over time. By providing empirical evidence on the prevalence of discussions about prejudice in global media, this study aims to enhance our understanding of media reporting on prejudice, thereby contributing to the development of theoretical frameworks within the fields of sociology and communication studies.

# Methods

Countries were selected for inclusion in our analysis by prioritizing population size, a diversity of world regions representativeness, historical news data availability and geopolitical strategic importance. A minimum of two news outlets per country was used as a threshold for inclusion of a country in the analysis. The average number of news outlets analyzed per country was 3.6. News outlets were selected by readership volume, availability of historical content in their online web domains and technical feasibility of automated content analysis of their web domains. We prioritized news outlets with English content for our sample to facilitate the analysis, but we also analyzed 41 outlets in other languages such as Spanish, French, German, Swedish, Italian or Portuguese. When comprehensive English versions of the news outlet content were available in the news outlet web domain, we used that data in the analysis. We translated our set of target words in English (*sexism*, *racism*, *diversity*, etc.) to the different languages of news outlets analyzed using Google Translate.

The text content of the analyzed articles is available on the web domains of the respective media outlets and often also in Internet cache repositories such as Common Crawl, Google cache or the Internet Archive. We didn't use any proprietary data sets in our analysis. We provide a reproducibility data set of counts of target terms and total number of unigrams in each article plus a Google searchable prefix of the headline of the article in order for others to verify the integrity of the data set and the frequency counts. We note that some news article content is occasionally behind pay walls. For some of those instances, the full text of the article is still accessible in the page source. For cases where this is not the case, we often relied on just the open access version of the article which does not contain the full article text. Yet, the open access version still usually contains the first few paragraphs of the article and the headline, thus still serving as a useful proxy to approximate prevalence of terms in the news outlet content.

Our content analysis is limited to the headlines and text body of the articles and does not include other article elements such as figure captions or subheadings. This work has not analyzed the audiovisual content of media outlets, except when said outlets offer a transcription of their audiovisual content in written form. The text of the articles was located in the HTML source code of each article URL using outlet-specific XPath expressions. All tokens (unigrams) in the corpus were converted to lowercase letters before frequency estimation.

The annual relative frequency $f_{io}^y$ of a target term $i$ (i.e. a unigram like *diversity* or n-gram like *social justice*) in a given outlet $o$ in a given year $y$ was estimated by dividing the number of occurrences of the target term $i$ in all outlet $o$ articles from year $y$ by the total number of words (i.e. unigrams) in all outlet $o$ articles for year $y$. This relative frequency metric is robust to the variable number of articles and text volume published by media outlets in different years. A sequence of relative frequencies of a term $i$ in an outlet $o$ over several years, constitute the time series ($f_{io}$) of relative frequencies for that term in the outlet.

$$f_{io}^y = \frac{number\ of\ mentions\ of\ ngram\ i\ in\ outlet\ o\ in\ year\ y}{total\ number\ of\ unigrams\ in\ outlet\ o\ in\ year\ y}$$

In order to avoid content from an outlet with a low volume of text in a given year producing noisy frequencies estimations, we established a minimum threshold for inclusion of outlet content from any given year consisting of the existence of at least 250,000 unigrams of content per outlet and year.

The temporal availability of articles in different media is not uniform. Some outlets have archived news content since the year 2000 and earlier in their online web domains. Others only have availability of content for more recent years, see Supporting Information (SI) for a breakdown of article volume analyzed per outlet and year.

For a small percentage of articles, our analysis scripts might not be perfectly accurate at locating the article's headline and body of text in the article's source code. This is due to the heterogeneity of HTML code and CSS styles in which articles' text is embedded in the web pages of the analyzed media.

Most of the erroneous frequencies are minor deviations from the correct frequencies such as counting a target word in a figure caption and which our analysis scripts incorrectly included as belonging to the main body of the article. In short, in a data analysis of more than 98 million articles, it is not feasible to manually count the number of mentions of a target term in all articles. Minor errors may occasionally occur in the automated computation of terms' frequencies due to factors such as incorrect HTML source code or partial article content behind a pay wall. But in general, our method of estimating word frequencies in media content produces verifiable results as illustrated by our previous work [1], [3], [4], [13] and Figure 2 that shows the relative frequencies of several illustrative terms in New York Times articles over the last 50 years. The figure shows how our method accurately captures the temporal dynamics of the first Gulf War (1991), the Iraq War (2003), the presidencies of Bush Senior and Junior, the ascendance of China as a global superpower, the disappearance of the Soviet Union, the waves of technologies that come and go (cassettes, DVDs, Netflix), the rise or fall in prominence of corporations (Facebook vs General Motors), the various editions of the Olympic Games and other social/natural/economic phenomena (i.e. anxiety, climate change, bitcoin, etc.).

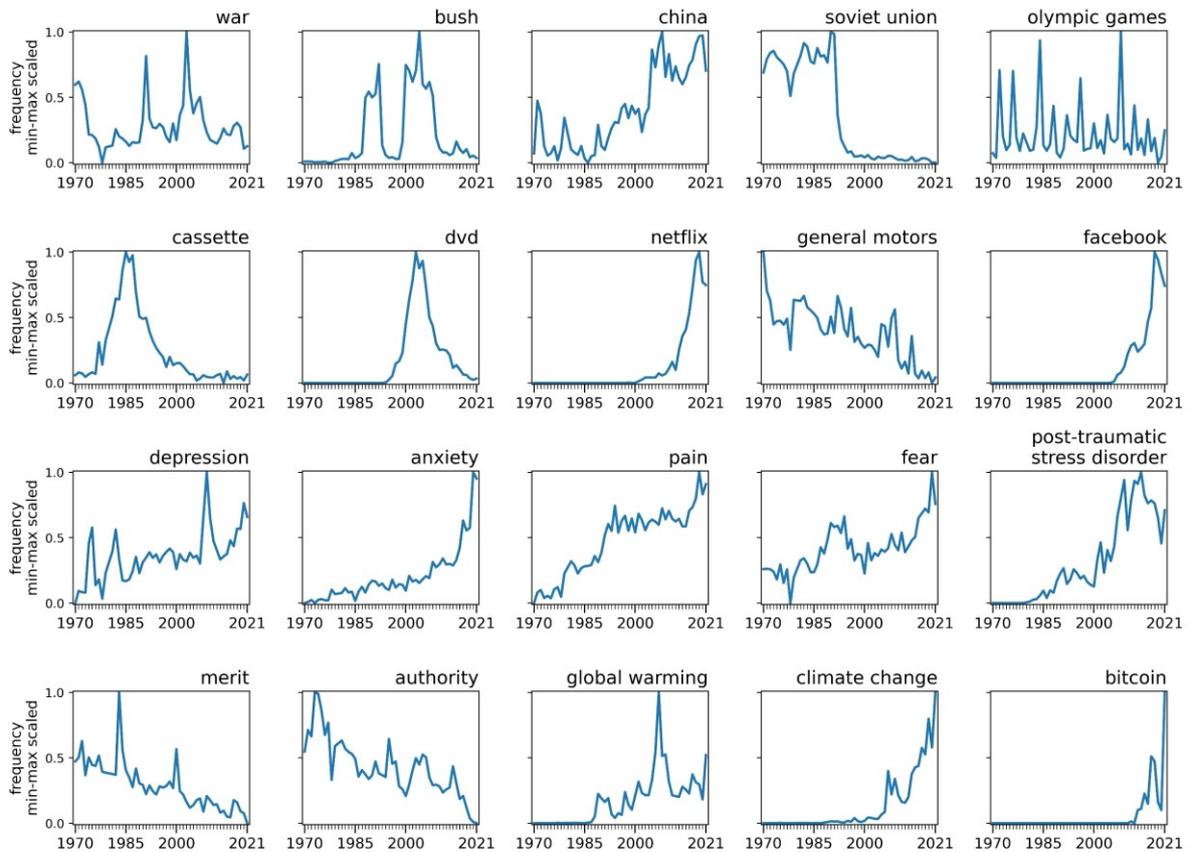

*Figure 1 Min-max scaled yearly frequency of word usage in New York Times articles for a set of illustrative terms.*

**Averages of min-max scaled relative frequencies time series**

Analyzing the prevalence of a theme in a corpus of news media articles requires the usage of several vantage points to gather a holistic picture that fully characterizes the phenomena. When attempting to quantify the chronological prevalence of a construct such as *prejudice-denouncing terms* (i.e. *racism*, *sexism*, *homophobia*, *transphobia*, *Islamophobia*, etc.), simply aggregating the relative frequencies of each term by summation or averaging is not advisable since the aggregated metric will be dominated by any term in the construct with an outsized frequency magnitude in comparison to the other terms.

That is, if the word *racism* is used 10 times more often in the corpus than the word *islamophobia*, adding the time series' frequencies of each term or taking the average will produce a distorted view of the construct that will be dominated by the highest frequency term while obscuring the lower prevalence terms dynamics. Using other central tendency estimations that mitigate the impact of outliers, such as harmonic or geometric means of the time series, is unfeasible in our data set since many outlets have zero frequency values for certain terms in some years.

We therefore apply *min-max feature scaling* to bring each term relative frequencies time series to the range 0-1, (denoted as $f'^{y}_{io}$), so as to neutralize the impact of higher frequency terms on the aggregated dynamics of the construct.

$$f'^{y}_{io} = \frac{f^{y}_{io} - \min(f_{io})}{\max(f_{io}) - \min(f_{io})}$$

Our study analyzes constructs of terms (i.e. prejudice denoting terms or social justice related terminology). We take the average of all the min-max frequencies in the construct and min-max the resulting average to rescale to the 0-1 range. This approach produces similar dynamics to factor analysis, but it is more transparent and does not need arbitrary hyper-parameter tuning like factor analysis does for eigenvalue cut-offs, rotation method to use, number of factors to select, thresholds for inclusion of a word in a factor, etc. The resulting average min-max scaled frequencies metric time series provides a good assessment of overall dynamics of the terms in the construct that allows discerning periods of minimum, medium and maximum usage of the terms in the construct overall. Of course, this metric represents just one vantage point of the phenomena we are analyzing, so we complement this analysis with others like relative frequencies, breaking the analysis by prejudice type, and relative frequencies rate of change.

## Results

We first analyze average min-max scaled frequencies of terms commonly used to denounce prejudice (i.e. *racism*, *sexism*, *homophobia*, etc.) in 124 popular news media outlets from 36 countries representing six different world regions, see Figure 2. The figure shows that in most of the countries analyzed, there has been an increase in news media mentions of prejudice-denouncing terms post-2010. Note that in order to make the trends clearer, we have applied a three-year moving average smoothing function to the plots.

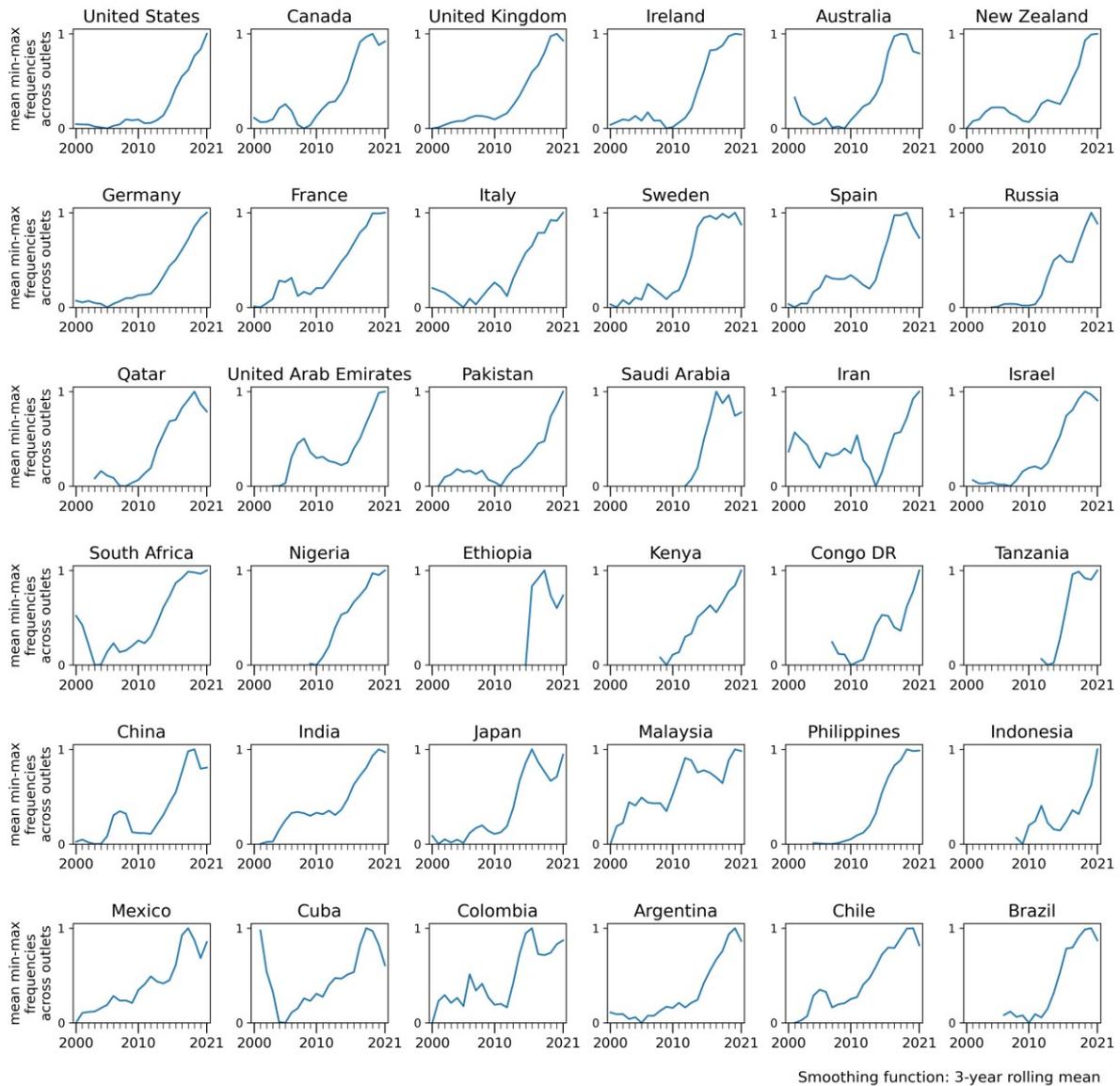

*Figure 2 Average min-max scaled yearly frequencies of prejudice-denoting terms in popular news media outlets across 36 countries representing 6 different world regions. From top to bottom rows: English-speaking West, continental Europe, Persian Gulf region, sub-Saharan Africa, Asia and Latin America.*

Figure 2 aggregates all the words mentioning different prejudice types into a single aggregate metric, thereby obscuring prejudice-specific dynamics globally as well as prevalence of prejudice-specific themes by country. To study prejudice-specific dynamics worldwide, Figure 3 shows the average global relative frequency across the different prejudice types studied. All prejudice types have, on average, increased globally in prominence since at least 2010. The dynamics of the individual terms in Figure 2 are shown as SI.

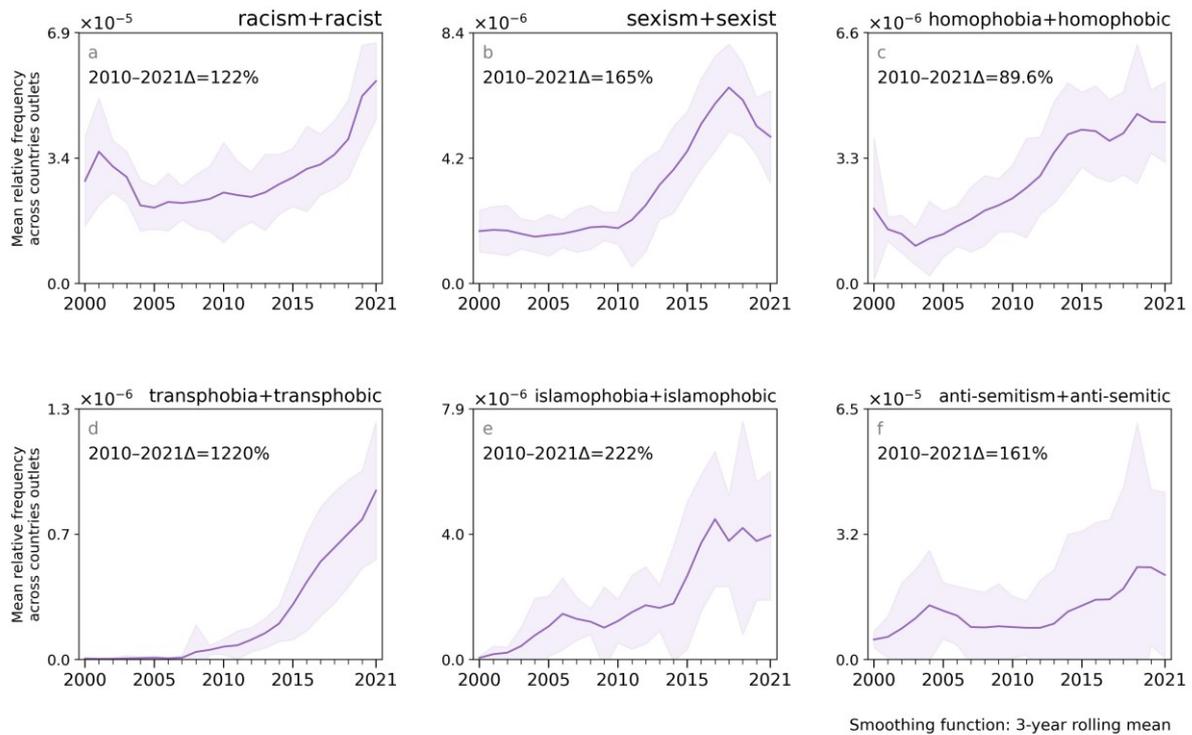

*Figure 3 Average prevalence of prejudice-denoting terms in 124 popular news media outlets from 36 different countries. The shaded areas around the mean trends display the 95% confidence intervals. The 2010 to 2021 percentage change in frequency is shown on the upper left of each subplot.*

To study prejudice-specific relative frequencies in individual countries, Figure 4 lists the average relative frequencies within the 2015–2021 time interval of different prejudice types in countries' news media. It is clear from the figure that the topic of racial prejudice is very prevalent in English-speaking Western countries, continental Europe and some Persian Gulf region countries like Israel and Qatar. Prevalence of this topic is lowest in sub-Saharan African (with the exception of South Africa) and Latin American countries (with the exception of Brazil).

The sexism prejudice topic manifests a linguistic quirk by which some languages prioritize some words over others to refer to it. For instance, in English-speaking countries the term *sexism* or *sexist* are the main terms used to refer to gender prejudice while *machismo* or *male chauvinist* are secondary. In Spanish speaking countries it is the exact reverse order. That is, *machismo* and *machista* are the more prominent terms to refer to gender prejudice while *sexismo* and *sexista* are less prevalent. Thus, for this prejudice type we used a composite of terms that signify gender prejudice in order to properly measure the overall prevalence of this topic in news media content from different countries.

Figure 4 shows that the topic of gender prejudice is most prevalent in the West (both English-speaking countries and continental Europe) and Latin America. The topic is least prevalent in several Gulf region countries, Asian and African countries. The topic of gender prejudice is particularly prominent in Spanish news media, with an average prevalence that almost triples in size the prevalence of this topic in some of the other countries that manifest the largest prevalence of the gender prejudice theme, France and the United Kingdom.

The topic of sexual orientation prejudice is very prevalent in continental Europe, English-speaking West, Israel, South Africa and Latin America and less mentioned in most Gulf region, Asian and African countries.

The topic of gender identity prejudice displays very similar dynamics to the topic of sexual orientation prejudice, being most prevalent in continental Europe, English-speaking West and Latin America and much less mentioned in most Gulf region, Asian and African countries.

The topic of Islamophobia is most prevalent in Gulf region countries (including Israel), France, Canada and the U.K. as well as some Asian countries with a large Muslim population such as Indonesia and Malaysia. This topic is much less prevalent in African and Latin American countries.

The topic of anti-Semitism displays an outsized prominence in Israeli news media. The other countries that display the largest prevalence of this topic in their news media content (albeit much less than Israel) are mostly Western countries like Germany, France, the UK or the United States and some Gulf region countries like Saudi Arabia, Qatar and Iran. The topic is barely mentioned in Asian, Latin American and African countries news media.

Some overall interesting patterns in Figure 4 are that news media from Western countries, Russia and some gulf region countries display the heaviest usages of prejudice-denouncing terminology in news media content. Latin America news media shows similarities to their Western peers in the prominence it assigns to the topics of gender, sexual orientation and gender identity prejudice but it dedicates much less attention to the topics of ethnic prejudice or religious orientation prejudice. The United States news media uses references to prejudice above the worldwide average, but it is not the heaviest user of prejudice-denouncing terminology among its Western peers. South Africa news media is an outlier in terms of using references to prejudice in their content relatively often in comparison to other African countries. Israeli news media also heavily uses prejudice-denouncing terms in their content in comparison to other countries in the Gulf region. The intensity of anti-Semitism mentions in Israeli news media is over 10 times larger than mentions of islamophobia in news media from predominantly Muslim countries in the Gulf region such as Iran, Qatar, United Arab Emirates, Pakistan or Saudi Arabia. Mentions of anti-Semitism in Israeli news media is more than four times larger than mentions of racism in American news media.

We complement the preceding analysis in the Supplementary Information with the rates of change in the usage of prejudice denoting terminology broken by prejudice types and country.

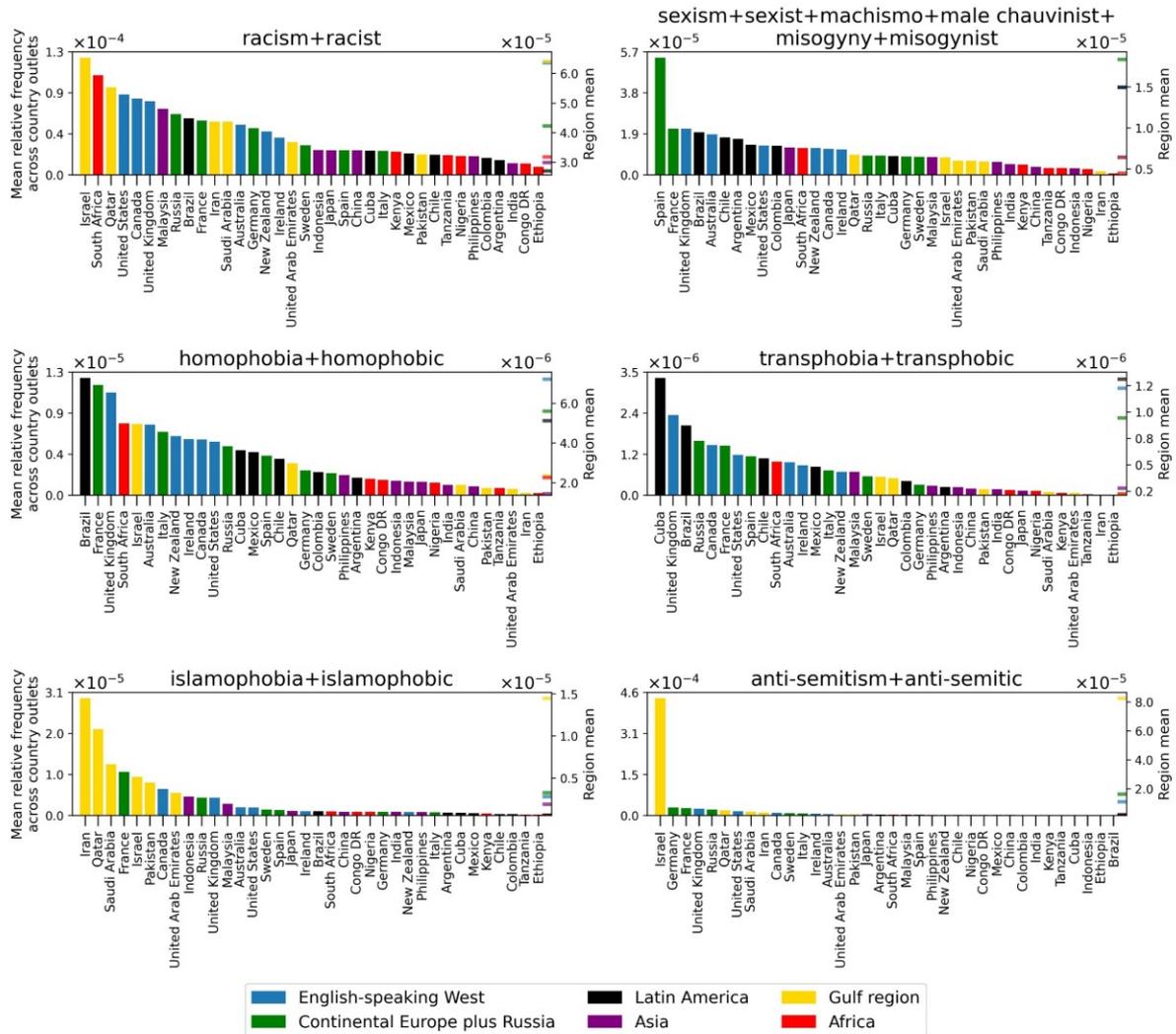

*Figure 4 Average relative frequency of terms that denounce prejudice in countries' news media content within the 2015–2021 time interval.*

To assess the temporal dynamics of how the increasing prevalence of prejudice denoting terminology has come about in different world regions, Figure 5A shows the average min-max scaled frequencies of our sample prejudice-denoting terms in the six different world regions studied. The trend appears mostly moderately in sync across the six world regions studied. A discrete differentiation of the regional time series is provided as SI and shows that the regional time series maximum rate of growth occurred simultaneously in Africa, English-speaking West and Gulf region in 2016, a year earlier in Latin America and Asia (i.e. 2015) and two years earlier (i.e. 2014) in continental Europe.

In Figure 5B we plot some of the countries that appear to peak first in their usage of prejudice signifying terminology plus the United States. The figure suggests that American news media, on average, was not first in increasing deployment of references to terms that signify prejudice and that news media from countries such as Sweden, Colombia or Canada began to increase the deployment of prejudice denouncing terminology earlier than news media from the United States (see SI for a visualization of the differenced time series). Similarly, when plotting some international news media outlets and the newspaper of record in the United States, the *New York Times*, we can also observe that the *New York Times* was not first in increasing the usage of prejudice signifying terminology in its content.

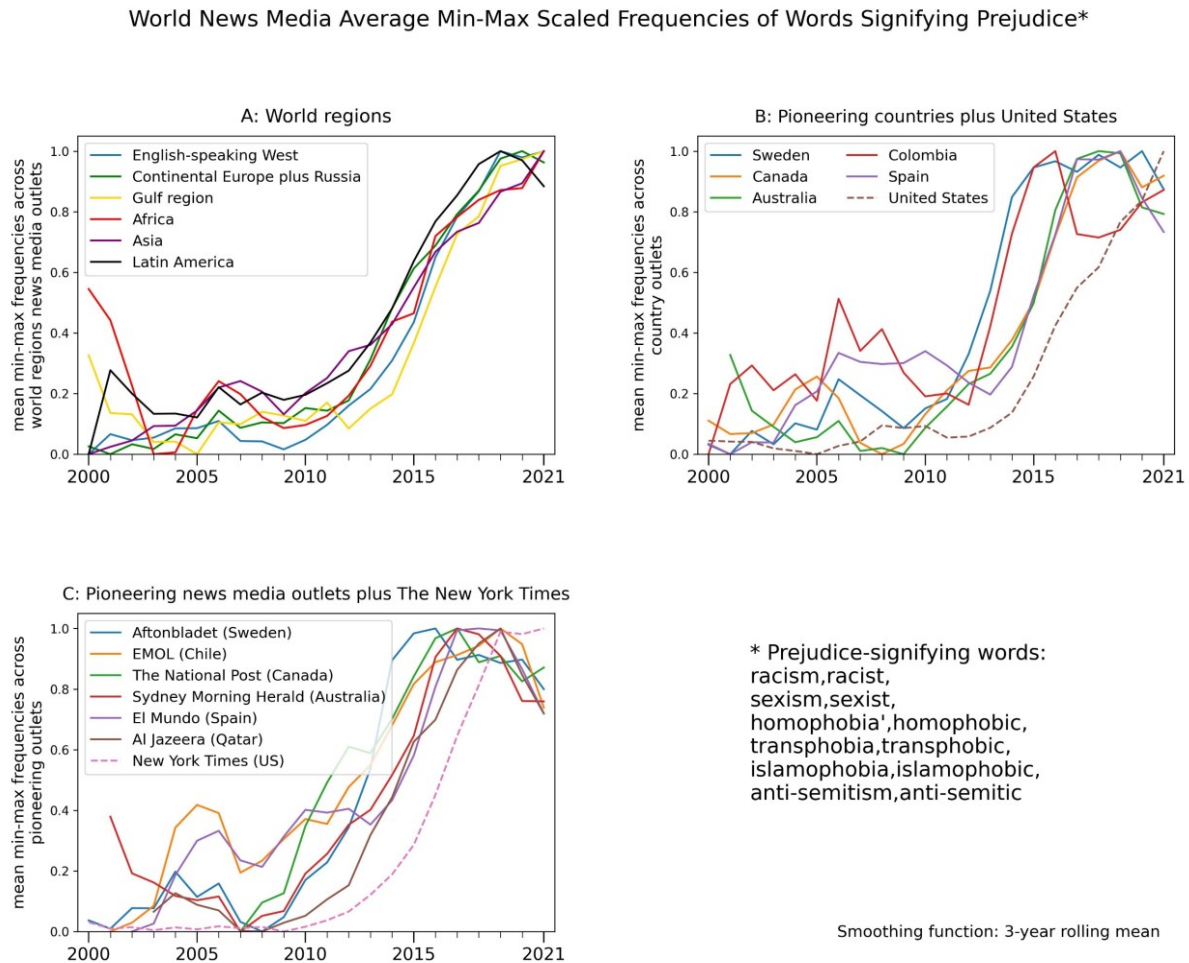

*Figure 5 A: Average min-max scaled yearly frequencies of terms used to refer to prejudice in popular news media outlets across six different world regions. B: Average min-max scaled yearly frequencies of terms used to refer to prejudice in several countries plus the United States. C: Average min-max scaled yearly frequencies of terms used to refer to prejudice in several popular news media outlets from different countries.*

To conclude our analysis, we plot the average min-max scaled frequencies of an additional set of terms often associated with social justice discourse and positive connotations such as *diversity*, *equity, inclusion* or *fairness*, see green trend in Figure 6. Note that in order to make the trends clearer we have applied a three-year moving average smoothing function to the plots. These terms also display a marked increase in prevalence post-2010. Furthermore, their dynamics are highly correlated with the previously studied prejudice-denouncing terms in most of the studied countries (see the large Pearson correlation coefficient, r, between the green continuous trend tracking the prevalence of terminology often associated with positive social justice discourse and the dashed blue trend tracking the usage of prejudice-denoting terminology loaded with negative connotations).

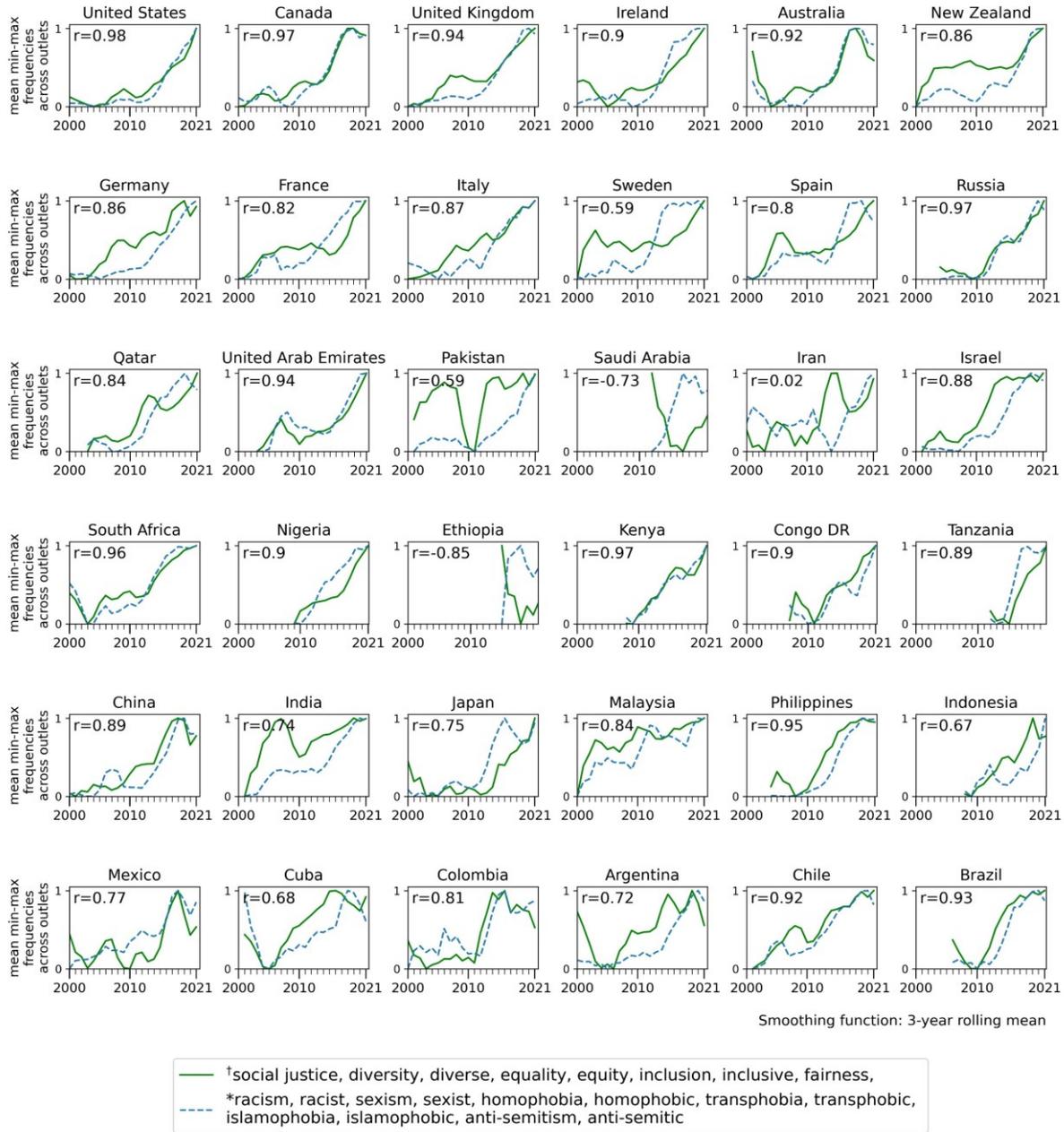

*Figure 6 Average min-max scaled yearly frequencies of positive terms often associated with social justice discourse (green continuous time series) and negative terms used to refer to prejudice (blue discontinuous time series) in popular news media outlets across 36 countries representing six different world regions. From top to bottom rows: English-speaking West, continental Europe, Persian Gulf region countries plus Israel, sub-Saharan Africa, Asia and Latin America. The Pearson correlation coefficient between both time series is shown on the upper left of each plot.*

The previous analysis presents a picture of global news media increasing usage of prejudice denouncing terminology and social justice discourse from 2010 to 2021. However, are all countries using these lexicons for the same purpose or with the same intention? A perfunctory qualitative analysis of Iran's and China's state media such as the *Tehran Times, Islamic Republic News Agency, People's Daily or China Daily* suggests that this is not the case (see SI for details). While in Western news media, most mentions of prejudice occur in a context of denouncing prejudice, mostly within Western nations, mentions of terms signifying prejudice in China and Iran are almost never used introspectively to denounce prejudice within their frontiers but rather they appear to mostly be used to highlight alleged prejudice in the West. In the case of China and Iran state media, it is specifically alleged ethnic/racial prejudice in the West that is mostly mentioned.

## Discussion

This work has documented the increasing prominence of prejudice denouncing terminology in news media worldwide starting post-2010 but prior to 2015 in pioneering countries. An important question emerging from our results is to wonder about the reasons why news media worldwide are increasingly using terms that denote prejudice in their content. Next, we outline several hypotheses to suggest potential contributors to this trend.

A potential reason for the media phenomenon traced in this work could be that prejudice itself is increasing worldwide and news media is just reporting about it. Although conceivable, this hypothesis would be paradoxical by representing a reversal of a decades-long decreasing trend in explicit prejudicial attitudes that started in the 1970s, at least in the U.S. [14], [15], [16], [17], [18], [19], [19], [20].

Alternatively, countries worldwide might have become more adept at noticing and denouncing prejudice against groups that have been prescribed as protected. Others have argued that a relaxation of the criteria used to define what constitutes prejudice, i.e. *concept creep*, could be partially responsible for this phenomenon [21], [22]. A mediator for this trend could be a cognitive bias known as *prevalence induced concept change* which has been shown experimentally to reduce the threshold used to establish membership of a category for a phenomenon as the frequency of said phenomenon decreases [23]. Both higher accuracy at detecting and denouncing prejudice as well as a decrease of the threshold used to established what counts as prejudice would be consistent with the increasing frequency of prejudice denouncing terminology in news media even if prejudicial attitudes in society were decreasing or stable.

Instead, news media worldwide could simply be mimicking trends in United States news media regardless of prejudicial attitudes trends in their own nations due to the presumed oversized cultural influence of the United States on other countries. This hypothesis however is partially incompatible with the fact that the United States news media does not appear to have been the pioneer in embracing the increasing usage of prejudice denoting terminology in their content. That is, other countries' news media appear to have started increasing their usage of prejudice denoting terminology before the United States.

Another possibility to explain the trends described in our results could be cultural changes such as the hypothesized emergence of *victimhood culture* that allegedly incentivizes the cultivation of victimhood identity due to the purportedly sympathy bestowed upon alleged victims of social ills [24], [25], [26]. According to this hypothesis, in situations of competition or social conflict, individuals and groups would have incentives to claim or exaggerate an identity of victimhood as a mechanism to obtain a competitive advantage in the conflict by appealing to social or institutional sympathy. If true, such an underlying societal trend could also partially explain the increasing prevalence of prejudice-signifying terminology in news media.

Relatedly, societies that actively oppose victimization of groups prescribed as protected might also confer status on those that protect or vindicate victims. Other authors have studied how public expressions of moral virtue (*moral grandstanding*) like strident opposition to social ills, such as prejudice, could not only be used in an altruistic manner but also as a mechanism to signal moral purity and thus gain social prestige [27]. If these hypotheses were true, such psychosocial processes could also partially explain the increasing frequency of terms that denounce prejudice in news media.

Another hypothesis to explain the media trends described in this work could be the increasing intellectual homogeneity in newsrooms. The lack of ideological diversity within the journalism profession has been described before in the United States [28], the U.K. [29] and 15 others Western countries [30]. That is, most journalists' political orientation sits left-of-center in comparison to the general population. Furthermore, the political orientation homogeneity among journalists appears to be increasing over time [28], [29]. Other studies have also shown that concern with prejudice is particularly high among people who manifest left-leaning political orientation[31], [32]. Therefore, an increasing proportion of journalists' professionals leaning left-of-center politically could cause growing media coverage of topics such as prejudice since preoccupation with prejudice is particularly marked among left-leaning oriented individuals.

Yet another potential explanation for the rising incidence of prejudice and social justice rhetoric in news media content could be the recent emergence of financial incentives for media organizations to maximize diffusion of news articles through social media channels by triggering negative sentiment/emotions, and/or political out-group animosity, both of which have been shown to drive engagement of social media-based news consumption [33], [34]. Indeed it has been shown previously that negative affect in news media content is increasing [13], [35], [36]. Because prejudice denoting terms such as *racism*, *sexism* or *homophobia* convey a very negative emotional payload, such terms could be being leveraged in headlines to maximize click-through ratios in both social media channels and outlets' institutional web domains.

One additional hypothesis that could explain the patterns described in this work would be that such trends are not organically emergent but that instead they are the result of purposeful guidance/coordination. Potential mechanisms for this hypothesis would be the relatively small number of media conglomerates that own many of the most popular news media outlets and news agencies, conceivably affording them influence over editorial content if they would choose to do so. Relatedly, news media outlets worldwide have experienced in recent years financial woes as a result of declining advertising revenue and newspaper sales. As a result, such organizations are more susceptible to receive funding from institutional or private donors, as has happened occasionally [37], [38], [39]. It is conceivable that at least some of those donations might not be exclusively altruistic in nature and could be seeking to influence media content. Whether those donations have an impact on content is obviously an open question. Here we are just listing the myriad of possibilities that could explain the underlying phenomena we have described in this work.

All the previously listed hypotheses are at least conceivable. However, it is highly unlikely that they all apply equally to all the countries analyzed with their vastly different historical and social contexts. It could well be that the increasing prominence of prejudice denoting terms in news media worldwide could be the result of different causes depending on the social context of each country. The phenomenon nonetheless could still reach synchronicity worldwide because of the interdependence and connectivity derived from globalization.

Thus, some Western countries could be increasing the prominence of prejudice denouncing terminology in their content by a combination of factors such as genuine concern for the existence of prejudice in society, increasing ideological homogeneity in newsrooms, *prevalence induced concept change*, *victimhood culture*, *moral grandstanding* and donors interested in raising awareness about prejudice. Journalists in other countries that perhaps assign moral authority to Western liberal democracies, could be following in close temporal proximity the editorial trends established by Western media. Yet, other countries, especially adversarial ones, could be leveraging prejudice-denouncing terminology in their state-controlled news media as a mechanism to criticize or denounce Western nations. This multifactorial scenario could still explain the synchronicity of the pattern worldwide.

Our preliminary qualitative analysis indeed hints at how some outlets from China and Iran use prejudice denouncing terminology not to introspectively criticize their own countries/societies but instead to point out prejudice in Western nations in general and the U.S. in particular. Media from both countries tend to focus mostly on ethnic prejudice not on other types of prejudice. Russia's state-controlled RT instead substantially uses prejudice-denoting vocabulary, but often as a tool to mock what they seem to perceive at the excesses of the West with respect to so-called social justice. We emphasize however that our qualitative analysis is markedly preliminary and insufficient in scope. More in-depth qualitative analyses about how much reporting in countries beyond the U.S. is focused on commenting on developments in the U.S. is necessary.

The strong association in news media content between prejudice signifying words and social justice terminology with positive connotations is noteworthy. The extremely high correlation between both sets of terms in news media across different countries suggests that they are related to the same underlying phenomena. Yet their very different emotional valence (extremely negative in the case of prejudice signifying words and markedly positive in the case of positive social justice rhetoric) suggests the multifaceted nature of the so-called *Great Awokening*.

An important follow-up of this work would be to establish the impact on public opinion worldwide of the increasing prominence of prejudice-denoting terms in news media. Previous academic literature has demonstrated that news media can play an important role in setting the agenda of priorities in public opinion. This phenomenon has been named *agenda-setting*. It has been shown before also that when news media increases its coverage of phenomena such as terrorism or crime, public opinion preoccupation with respect to those topics increases, irrespective of actual levels of terrorism or crime in society [40], [41]. Therefore, the increasing prevalence of prejudice signifying terminology in news media content could be causing increasing public concern about the severity of prejudice in society, as predicted by *agenda-setting* theory. There is some limited evidence that this could be partially happening in the U.S., but more empirical work is needed on this topic [1].

Our results also seem to indicate that some prejudice themes are no longer increasing or have even slightly decreased in prevalence post-2020 in several of the analyzed countries. This could be due to societies or newsrooms having reached peak-concern with some types of prejudice between 2015-2019. Alternatively, the beginning of the COVID-19 pandemic in 2020 could have grabbed such a substantial amount of news media attention as to distort the broader trend, which may well resume as the global pandemic recedes. The end of the controversial Trump presidency in the U.S. could also be playing a role. A potential return of Trump to the White House could potentially reignite the upwards trend.

An important limitation of our methodology is that the frequency metric used in our analyses lacks critical information about the context in which prejudice-denoting terms are being used. Qualitative analysis techniques are needed to properly establish the semantic loading with which the terms are being used. We have carried out such a perfunctory preliminary analysis to contrast the usage of prejudice denoting terminology between Western news media and critics of the West such as China, Russia or Iran. But more exhaustive work is needed on this front.

A further limitation of our methodology is that the construct of terms that we used to track the topic of prejudice in news media is limiting as it is based on lexical search. That is, it does not include alternative formulations of the concepts outside the list of target terms analyzed. For instance, in the supplementary material we include a plot about the frequency with which the terms *islamophobia* and *islamophobic* were used by U.S. media post-2001, a period that included the aftermath of 9/11 and the wars in Afghanistan and Iraq. Yet there is only a marginal increase in mentions of *islamophobia* and *islamophobic* in U.S. media in the first decade of the 21$^{st}$ century. The most likely explanation for this paradox is that the usage of *islamophobia* and *islamophobic* is probably fairly recent and between 2000 and 2010 other terms such as "bigotry towards…," "hatred for…," "discrimination against…," or "intolerance of…" were the preferred option used to refer to hatred towards minorities such as Muslims. More advanced and newer search techniques such as semantic search that locate target terms *and* their synonyms in semantic space could address this limitation and this is a potential direction of future work.

Another important limitation of this work is the variable news media sample representativeness for different countries. While some countries in our analysis like the United States, Colombia, Canada, Germany or Spain have a multitude of outlets in the sample, other countries have more limited sample sizes like for instance Japan which in our data set is only composed of two news outlets. Yet, our previous research hints that news media usage of terms that denote prejudice is highly correlated between outlets within a country [1], [2], [3]. So, perhaps this limitation is not too severe in terms of impacting the generalizability of the results. Additionally, while we often used news media content in the local language when possible, occasionally we had to settle for local news media with English content, like China's *People's Daily* or Japan's *The Asahi Shimbun*. It is conceivable that there might be differences between those outlets content in English and the same outlets content in the country's local language since they target different audiences. Yet, the English content versions still provides a valuable, if yet imperfect, proxy to gather insight about a country news media content.

There are some forms of prejudice that we have not explicitly covered in this work. For example, anti-immigrant and xenophobic sentiment which have been central features of election campaigns in Western democracies over the last years. To an extent, this topic is partially covered by the analysis of ethnic prejudice in this work. But anti-immigrant sentiment is such a vast topic that we feel it probably deserves a focused independent study of its own.

A final limitation of our analysis to establish the propagation of news media trends worldwide is our usage of yearly frequencies to investigate news media content. Finer granularity high frequency analysis at the monthly, weekly or daily rates could be illuminating to more precisely establish how news media trends propagate globally.

To conclude, we have documented a worldwide post-2010 substantial increase in news media of references to terms that denounce prejudice. Future research should try to elucidate the causal factors responsible for this global increase in news media usage of terminology that references prejudice. Despite the potential for a multitude of causes driving the trend, the synchronicity of the phenomena worldwide is noteworthy and deserving of future investigation. Of critical importance is to determine the impact on society of a public opinion that regularly consumes an information diet that increasingly contains references to the existence of prejudice in society.

# Conflict of interest statement

NA

# Data Availability Statement

The list of the over 98 million articles analyzed in this work and the frequency counters of the target terms and total unigrams per article are available in the following repository with DOI: https://doi.org/10.5281/zenodo.10983612